\documentclass[apj,twocolumn]{openjournal}

\usepackage{color}
\usepackage{comment}
\usepackage{natbib} 
\usepackage[dvipsnames]{xcolor}
\usepackage{aas_macros} 
\usepackage{amssymb}
\usepackage{amsmath}
\usepackage{threeparttable}
\usepackage[title]{appendix}
\usepackage{hyperref}	% Hyperlinks
\hypersetup{colorlinks=true,linkcolor=blue,citecolor=blue,filecolor=blue,urlcolor=blue}
\usepackage[caption=false]{subfig}
\usepackage{soul}   % provides \hl{} for highlighting

\defcitealias{zh3}{Z20}
\defcitealias{chandra}{C23}

\shorttitle{Magellanic Streams}
\shortauthors{}

\begin{document}

\title{Untangling Magellanic Streams}

%\author[0000-0002-5177-727X]{Dennis Zaritsky}
\author{Dennis Zaritsky$^{1}$}
%\author[0000-0002-0572-8012]{Vedant Chandra}
\author{Vedant Chandra$^{2}$}
%\author[0000-0002-1590-8551]{Charlie Conroy}
\author{Charlie Conroy$^{2}$}
%\author[0000-0002-7846-9787]{Ana Bonaca}
\author{Ana Bonaca$^{3}$}
%\author[0000-0002-1617-8917]{Phillip A. Cargile}
\author{Phillip A. Cargile$^{2}$}
%\author[0000-0003-3997-5705]{Rohan~P.~Naidu}
\author{Rohan~P.~Naidu$^{4,5}$}
\altaffiliation{$^5$NASA Hubble Fellow}

\affiliation{$^{1}$Steward Observatory, University of Arizona, 933 North Cherry Avenue, Tucson, AZ 85721-0065, USA}
\affiliation{$^{2}$Center for Astrophysics $|$ Harvard \& Smithsonian, 60 Garden Street, Cambridge, MA 02138, USA}
\affiliation{$^3$Carnegie Observatories, 813 Santa Barbara St., Pasadena, CA 91101, USA}
\affiliation{$^4$MIT Kavli Institute for Astrophysics and Space Research, 77 Massachusetts Ave., Cambridge, MA 02139, USA}

\email{corresponding email: dennis.zaritsky@gmail.com}

%%--------------------------------------- ABSTRACT ---------------------------------------%%

\begin{abstract}
\noindent
The Magellanic Stream (MS) has long been known to contain multiple  H{\small\ I} strands and corresponding stellar populations are beginning to be discovered.
Combining a sample of 17 stars from the H3 (“Hectochelle in the Halo at High
Resolution”) survey with 891 stars drawn from the {\it Gaia} DR3 catalog, we trace stars along a sub-dominant strand of the MS, as defined by gas content, across 30$^\circ$ on the sky. We find that the corresponding dominant strand at the similar position along the MS is devoid of stars with Galactocentric distance $\lesssim 55$ kpc while the subdominant strand shows a close correspondence to such stars. We conclude that (1) these two Stream strands  have different origins, (2) they are likely only close in projection, (3)  the subdominant strand is tidal in origin, and (4) the subdominant strand is composed of ``disk" material, gas and stars, with a chemical composition that marginally favors it coming from the Small Magellanic Cloud.

\end{abstract}


\section{Introduction}

The Magellanic Clouds are the nearest example of a phenomenon that we understand to be common throughout the history of the Universe --- the infall of small galaxies onto large dark matter halos \citep{blumenthal,davis}. As such, they present our best opportunity to study this central aspect of galaxy evolution and refine how it is modeled \citep[e.g.,][]{cole}.
Is gravity or hydrodynamics primarily responsible for stripping gas from such infalling galaxies \citep{mathewson,lin,moore}? How do galaxy interactions, either between the Clouds themselves or with the Milky Way (hereafter the MW), affect the star formation rates within the satellite galaxies \citep{z04,Harris_2009,massana}? How can such infall events affect the larger galaxy onto which the smaller galaxies have fallen \citep{weinberg,fox,gavarito,lucchini_21,carr}?

To address these and other questions, investigators have generated an increasingly sophisticated set of numerical simulations of the interaction \citep[e.g.,][]{lin,gardiner,besla_10,diaz,gomez,hammer,pardy,gavarito,wang,lucchini_21,lucchini_24,carr,jimenez}.  This progress has led  to a realization of the complexity of the system but also to a greater appreciation of its potential for leading us to a more complete understanding of such fundamental topics in astrophysics as the nature of the circumgalactic medium \citep{fox,lucchini_21,carr} and dark matter \citep{foote}.  
To motivate, constrain, and exploit even more complex simulations, an ever increasing set of detailed observational constraints is needed.

Among the variety of interesting features of the Magellanic system that challenge models is the long gaseous tail that trails the Clouds, referred to as the Magellanic Stream \citep[hereafter the MS;][]{dieter,wannier,mathewson}. The MS is a set of apparently intertwined H{\small\ I} strands \citep{cohen,morras,putman03,nidever_08} whose origin (LMC, Bridge, or SMC) is still debated \citep[e.g.,][]{nidever_08,fox,diaz,pardy}. The 3D geometry is unknown because of the difficulty in measuring distances to gas clouds \citep{putman03}. 

While arguments have often been framed around whether this gas was drawn out primarily by tidal forces \citep{fujimoto,lin} or ram pressure \citep{meurer,moore}, the unavoidable nature of tides, the weak, fragmented nature of the leading arm \citep{putman}, the ionized nature of the MS \citep{putman_03a}, and the morphology of high velocity clouds \citep{putman_11} suggest that both physical phenomena will be part of a full understanding. Potential internal factors, such as winds that push gas outward and make it more susceptible to either tides or ram pressure, add yet another layer of complexity  \citep{olano,nidever_08}. 

The tidal hypothesis for the origin of the MS offers hope for the presence of stars along the MS, which would then provide distance constraints, a better understanding of the geometry and, perhaps, of its origin. In contrast, the confirmed absence of stars in the MS would likewise be an important constraint and would favor a hydrodynamic origin scenario, such as the blowout plus stripping plus tidal model of \cite{nidever_08}. Searches for stars along the MS have generally yielded negative results \citep{recillas,brueck,raja}. More recently, tidal stellar features have been uncovered closer to the LMC-SMC system \citep{belokurov16,mackey,belokurov,deason,navarrete} but a population tracing the MS has been elusive. 

The H3 survey \citep{conroy} and follow-up observations have contributed to the body of work on the stellar counterpart of the MS in two studies. \cite{zh3}, hereafter \citetalias{zh3}, identified 15 stars that form a dynamically cold group that closely follows part of the MS in projection and matches roughly in radial velocity, but lacks the large negative angular momentum of the Clouds. \cite{chandra}, hereafter \citetalias{chandra}, identified 13 stars at greater Galactocentric distance, $60-120$~kpc, among a population of distant stars selected to match the large negative angular momentum of the Clouds about the Milky Way's x-axis. These, rather than the \citetalias{zh3} stars,  might appear to be more naturally identified as the stellar counterpart of the MS  because of the angular momentum match and the closer agreement in distance with the simulations of \cite{besla12}.  However, we advocate for caution in using models to validate interpretations given the sensitivity of the calculated Magellanic Cloud orbits to model assumptions \citep{vasiliev23} and the current state of sophisticated, but yet incomplete, hydrodynamical models  \citep{lucchini24a}.

If the \citetalias{zh3} stars are not part of the MS, then  what are they and what does the close association between these stars and at least some of the gas in the MS imply? The \citetalias{zh3} detection involved only 15 stars, making it difficult to trace the feature in detail and establish an unambiguous  association with any specific component of the MS. Here we expand the sample by using the \citetalias{zh3} stars to aid us in selecting a {\it Gaia} DR3 \citep{gaiadr3} sample of stars with which to better trace this population.
In \S\ref{sec:data} we describe how we select stars from the {\it Gaia} catalog. In \S\ref{sec:results} we describe the distribution of those stars and our inferences regarding the nature of the MS. In \S\ref{sec:discussion} we discuss certain aspects of our results and summarize in \S\ref{sec:conclusions}.

\section{Data}
\label{sec:data}

Our strategy is to use the H3 catalog to identify a pure sample of putative stellar stream stars, as identified by \citetalias{zh3}, and then use those stars to train our selection of additional possible members of this population in the {\it Gaia} catalog. We will then examine the resulting set of candidates to assess and interpret this population of stars and its association, if any, with the MS.

\subsection{H3 and the Selection of the Training Sample}

The H3 survey provides optical high-resolution spectroscopy of likely halo stars in a sparse grid covering roughly 15,000 square degrees \citep{conroy}. Likely halo stars are selected in high Galactic latitude fields  ($\vert b \vert>30^{\circ}$ and Dec. $>-20^{\circ}$), have an r-band magnitude between 15 and 18, and have a parallax that defines a lower distance bound ($ \varpi < 0.4$ mas). For further details, see \cite{conroy} and \cite{cargile2020}.

We obtain spectra of as many of these stars as we can using the fiber-fed Hectochelle spectrograph \citep{hecto} on the MMT in a configuration that produces spectra with a resolution of $\sim$ 32,000 from $5150$ to $5300$ \AA. 
From these spectra, H3 catalogs the stellar
parameters and spectro-photometric distances for $\approx$ 300,000 stars.
The procedure we use in determining stellar parameters and distance estimates was developed and presented by \cite{cargile2020}.
The values of $V_{GSR}$ are quite precise given that $V_{RAD}$ is measured to $\sim$ 0.5 km sec$^{-1}$ precision \citep[based on repeat measurements;][]{conroy} and the conversion to $V_{GSR}$ depends only on the position of the Sun in the Galaxy that we adopt.
From the available set of observed and analyzed H3 stars (rcat\_V4.0.5.d20240630\_MSG.fits)
we select those that have no spectral fitting problems (${\rm FLAG} = 0$), spectral signal-to-noise ratio (SNR) per pixel $>$ 2, are not identified as associated with the Sagittarius stream \citep[Sgr\_FLAG = 0;][]{johnson}, are not identified as being part of a cold kinematic structure (satellite galaxies or stellar clusters; coldstr = 0) and have a Galactocentric distance $>$ 30 kpc. The Sgr cut is an angular momentum cut corresponding to rejecting stars with $L_y < -2.5 - 0.3 L_z$, where the units are 10$^3$ kpc km s$^{-1}$\citep{johnson}.

We use the \citetalias{zh3} results to help guide a slightly revised kinematic selection. Originally, the selection was a straight cut on radial velocity. Here we select on energy, using only the radial velocity to evaluate the kinetic energy term, to help us find stars on nearly the same orbit across a range of distances. We adopt this approach in an attempt to detect stars both at smaller distances, where the crowding becomes more challenging, and at larger distances, where the radial velocity of any such stars might be 
significantly different than the original fixed cut. We set the energy threshold using the original set of stars and use galpy \citep{bovy} to calculate the radial
velocity as a function of distance in an NFW \citep{nfw} potential corresponding to a Milky Way with total mass, $M_{vir}$, $\sim 10^{12} M_\odot$ 
\citep{z89,shen} and then select stars within 30 km sec$^{-1}$ of that fiducial. As an acceptable simplification, the fiducial can be expressed as a third order polynomial: 
\begin{equation} \label{eq1}
\begin{split}
{\rm V}_{GSR} & = 4.042\times 10^{-5} {\rm R}_{GAL}^3 - 0.0178 {\rm R}_{GAL}^2 \\
&+ 2.87 {\rm R}_{GAL} - 295.1
\end{split}
\end{equation}
with R$_{GAL}$ expressed in kpc and V$_{GSR}$ in km s$^{-1}$.
There is some arbitrariness in the selection of the 30 km sec$^{-1}$ tolerance but it is meant to represent the potential velocity dispersion of stars in the original unknown progenitor. Different choices of this value yield larger or smaller samples but do not qualitatively affect the results we present.
Lastly, we limit the sample in the MS coordinate system \citep{nidever_08} to have  $-150^\circ < L_{MS} < -65^\circ$ and $-20^\circ < B_{MS} < 10^\circ$, which selects for stars in the vicinity of the MS.

\begin{figure}
\includegraphics[width=0.47\textwidth]{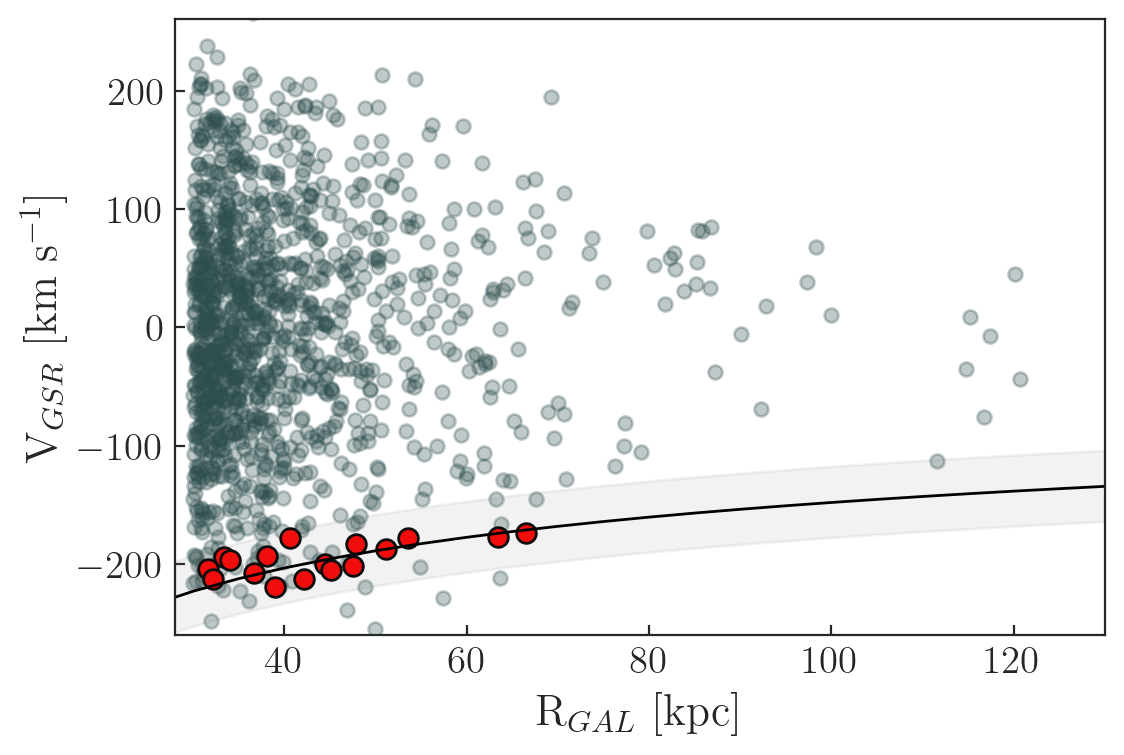}
\caption{Selection of candidate H3 stream stars. All H3 stars with R$_{GAL} > 30$ kpc, FLAG = 0, SNR $> 2$, Sgr\_FLAG = 0, and coldstr = 0 are plotted as gray circles. Our kinematic selection criteria is represented by the shaded area and encompasses stars on orbits of similar total energy in a 10$^{12}$ M$_\odot$ NFW potential that represents the MW. The 17 stars satisfying this and the other selection criteria described in the text are shown as red circles and comprise our H3-selected sample. Distance uncertainties in H3 are estimated at $\sim 10$\% and radial velocity uncertainties are smaller than or comparable to the plotted symbols.}
\label{fig:selection}
\end{figure}

\begin{figure}
\includegraphics[width=0.47\textwidth]{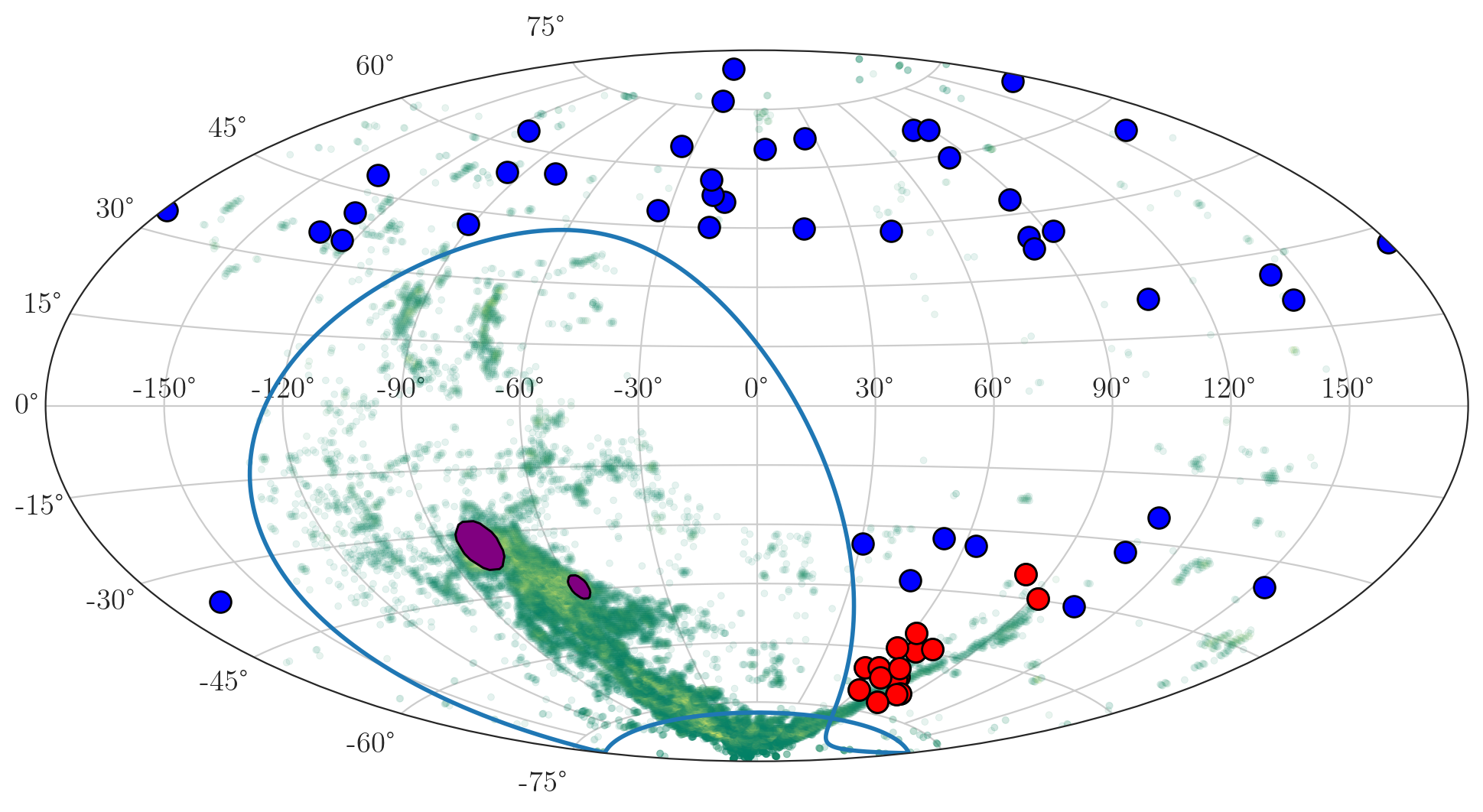}
\caption{All-sky distribution in Galactic coordinates of the gaseous Magellanic Stream \citep{nidever_08} in green, the LMC and SMC in purple ellipses, the H3-selected stars that satisfy all of our criteria in red, and those that satisfy all of our criteria except for the one related to the MS coordinate system in blue. The footprint of the H3 survey, which lies outside the enclosed region and avoids the Galactic disk, precludes us from finding stars along the MS that are closer to the Clouds than those shown. }
\label{fig:all-sky}
\end{figure}

Our selected set of stars is shown in Figures \ref{fig:selection} and \ref{fig:all-sky}. In Figure \ref{fig:selection}, the H3 stars that lie within the shaded region but are not selected are those that do not satisfy our MS coordinate system criterion. These stars are highlighted in blue in Figure \ref{fig:all-sky} and appear to be a random subsample of halo stars that happen to satisfy the kinematic criteria but do not lie along the MS. Note that in this context, the two stars near the tip of the MS are arguably part of the random halo sample rather than associated with the MS, but we retain them in our sample.

The sample of stars differs somewhat from the set of stars presented in \citetalias{zh3}, only 9 of 17 in the current set are in common.
This is primarily due to a change in Sgr\_FLAG that resulted in the rejection of 5 of the \citetalias{zh3} sample of 15. Two of these five stars fall near the tip of the MS. It is unclear if the addition of these stars helps to establish the population of stars associated with the MS tip or indicates that this population is better associated with Sgr. On the other hand, the remaining three stars fall well within the main group of stars and their inclusion or exclusion does not affect the interpretation of this population. We opt to continue to exclude the Sgr\_FLAG = 1 population but suspect that at least some of these stars may actually be part of the MS population we seek to identify.
The one other missing star from \citetalias{zh3} now falls about 20 km sec$^{-1}$ outside our revised kinematic criteria. Excluding these 6 stars does not affect our definition of the kinematic criterion. We will return to the topic of Sgr cross-contamination
in \S\ref{sec:results}.

\begin{deluxetable*}{llrrrrr}
\tablewidth{0pt}
\label{tab:stars}
\tablecaption{Selected H3 Stars}
\tablehead{  \colhead{H3 ID} & \colhead{GAIA EDR3 ID} &  \colhead{RA} & \colhead{Dec} & \colhead{R$_{GAL}$} &\colhead{$V_{GSR}$} & \colhead{SNR} \\
&&&& \colhead{[kpc]} &\colhead{[km s$^{-1}$]} & \colhead{[pixel$^{-1}$]}}
\startdata
113745468 & 2731996781084418432 & 341.7692179 &  13.314284 & 63 $\pm$ 6 & $-$177 $\pm$ 0.6 & 2.7\\
118022407 & 2812786554736110720 & 350.8647770 &  12.857316 & 33 $\pm$ 1 & $-$194 $\pm$ 0.4 & 4.5\\
118360373 & 2630501957940493184 & 348.3172866 &  $-$8.054611 & 51 $\pm$ 2 & $-$187 $\pm$ 0.1 & 13.0\\
119081183 & 2605618772853403776 & 345.1431990 & $-$11.273101 & 32 $\pm$ 3 & $-$204 $\pm$ 0.8 & 3.1\\
119163672 & 2633573237514082304 & 349.5990965 &  $-$5.356268 & 39 $\pm$ 6 & $-$219 $\pm$ 0.3 & 4.0\\
119511723 & 2610348528278657664 & 342.7162418 &  $-$7.813899 & 41 $\pm$ 1 & $-$178 $\pm$ 0.1 & 11.1\\
119716906 & 2433960715422572800 & 354.2856882 & $-$10.830486 & 67 $\pm$ 5 & $-$174 $\pm$ 0.4 & 3.9\\
119718763 & 2433738824527479168 & 354.4038651 & $-$11.362019 & 38 $\pm$ 2 & $-$193 $\pm$ 0.1 & 10.7\\
119719901 & 2433795548159999488 & 354.8523545 & $-$11.153222 & 44 $\pm$ 3 & $-$200 $\pm$ 0.2 & 6.1\\
119829926 & 2422797172003289088 & 359.9671116 & $-$10.730789 & 48 $\pm$ 2 & $-$201 $\pm$ 0.2 & 8.7\\
119830529 & 2422487281523040384 & 359.9326760 & $-$11.184418 & 48 $\pm$ 4 & $-$183 $\pm$ 0.3 & 5.2\\
120100224 & 2406304772463975808 & 348.1692334 & $-$16.703248 & 34 $\pm$ 2 & $-$196 $\pm$ 0.2 & 5.1\\
120116212 & 2408706415096654208 & 349.2970650 & $-$14.201853 & 32 $\pm$ 1 & $-$213 $\pm$ 0.1 & 16.4\\
120125054 & 2408253858687755264 & 352.4714532 & $-$13.831915 & 54 $\pm$ 5 & $-$177 $\pm$ 0.4 & 3.5\\
120442318 & 2436871844255646336 & 351.6534275 & $-$10.711871 & 45 $\pm$ 3 & $-$204 $\pm$ 0.3 & 4.4\\
120796507 & 2420734483893689984 & 000.4064159 & $-$13.861772 & 37 $\pm$ 4 & $-$208 $\pm$ 0.6 & 2.8\\
120984853 & 2394599028077116928 & 354.6068171 & $-$16.968077 & 42 $\pm$ 3 & $-$213 $\pm$ 0.4 & 4.3
\enddata
\end{deluxetable*}

The lack of H3 stars closer to the LMC and SMC along the MS is most likely due to the edge of the H3 survey footprint (see Figure \ref{fig:all-sky}). The dearth of stars farther along the MS reflects either a lack of such stars in reality or only in the catalog. The latter possibly because such stars are at greater distances than the H3-selected sample and so fainter and absent in the catalog. While our change in kinematic selection did not result in a large increase in sample size or radial range, it does now apply a more physically motivated velocity criterion. 

\subsection{{\it Gaia} Downselect}

\begin{figure}
\includegraphics[width=0.47\textwidth]{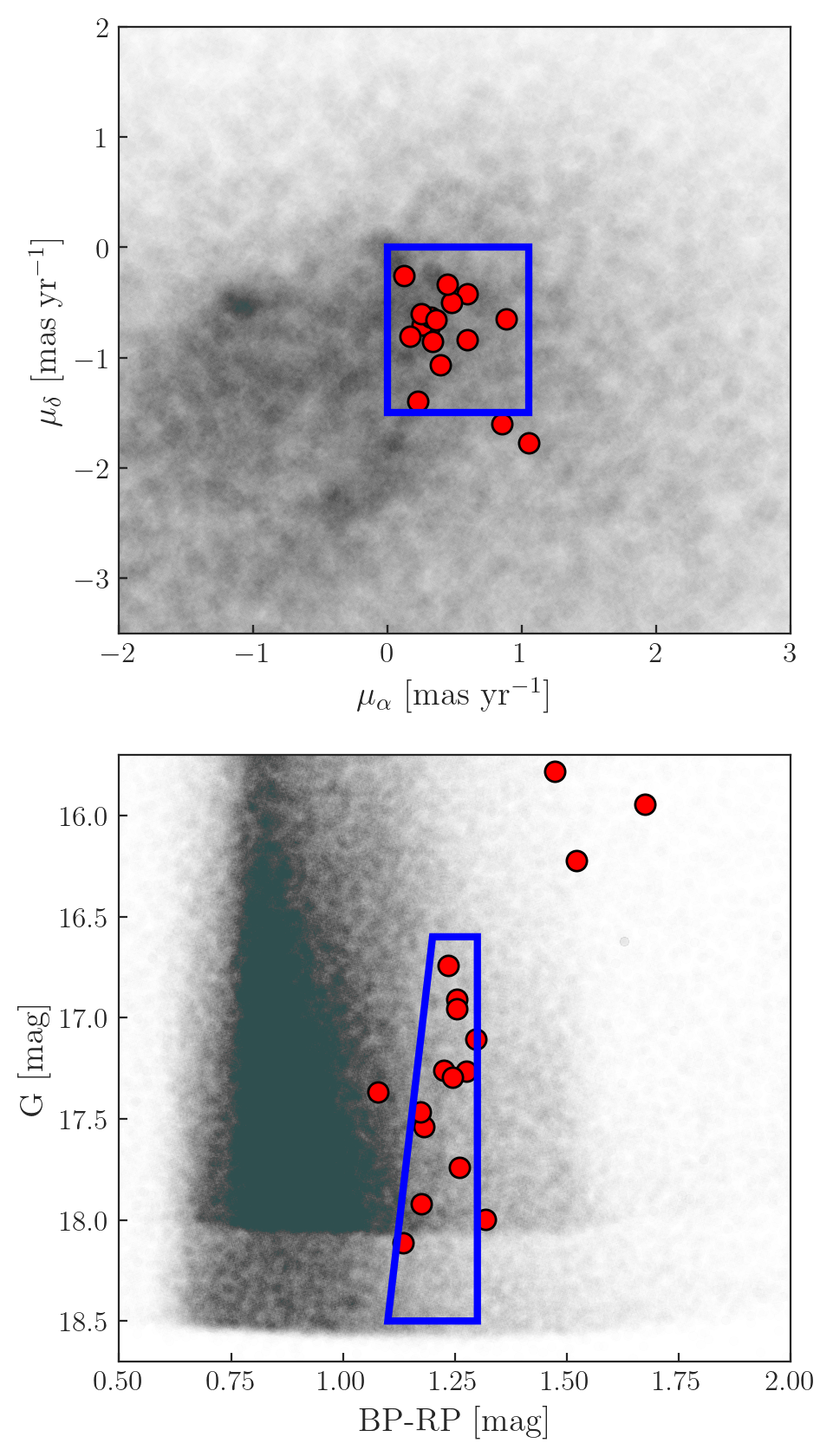}
\caption{The H3-selected stars (red circles) are localized relative to the full H3 sample in {\it Gaia} proper motion coordinates and color-magnitude space. Blue outlines mark our selection boxes.}
\label{fig:cuts}
\end{figure}

Our approach will work only if the properties of the H3-selected stars, which have precise radial velocity and distance estimates \citep[$\sigma_v \lesssim 1$ km sec$^{-1}$ and $\sigma_D/D \sim 10\%$;][]{conroy,cargile2020,chandra}, are sufficiently distinct within the parameter set measured by {\it Gaia} that does not include these measurements to enable us to select corresponding stars. 
Fortunately, our selected set of stars is both fairly consistent in its proper motion values and colors and distinct from the underlying population of stars (Figure \ref{fig:cuts}). First, we apply the basic parallax cut imposed in H3 ($ \varpi < 0.4$ mas), after applying a bias correction to the Gaia EDR3 parallax values \citep{gaia-correction}. Then, we interactively define our initial selection in proper motion to include most of the H3-selected stars: $0.0 < \mu_{\alpha}/({\rm mas\  yr})^{-1} < 1.05$ and $-1.5 < \mu_{\delta}/({\rm mas\  yr})^{-1} < 0.0$. We will trim more aggressively later but opt here to allow for some variation in proper motions among potential stream populations. In color-magnitude space we use the {\it Gaia} photometry and select stars with $16.6 < G < 18.5$ and $C < BP-RP < 1.3$, where $C \equiv (39.4-G)/19.0$. These cuts exclude the three brightest H3-selected stars, which also happen to be the reddest. These have colors and magnitude entirely consistent with being Magellanic Cloud stars \citep{gaia-mc}, but we concluded that enlarging the selection region to include these three added unnecessary complexity without producing significant return. Our aim is not completeness but rather a clean and minimally-contaminated sample.
Lastly, guided again by the H3-selected sample, we draw from the {\it Gaia} DR3 catalog \citep{gaiadr3} only stars with MS coordinates ($L_{MS},B_{MS}$) such that $-115^\circ < L_{MS} < -65^\circ$ and $-20^\circ < B_{MS} < 10^\circ$ to search for any existing population that closely tracks both the MS and the H3-selected stars.

\begin{figure}
\includegraphics[width=0.47\textwidth]{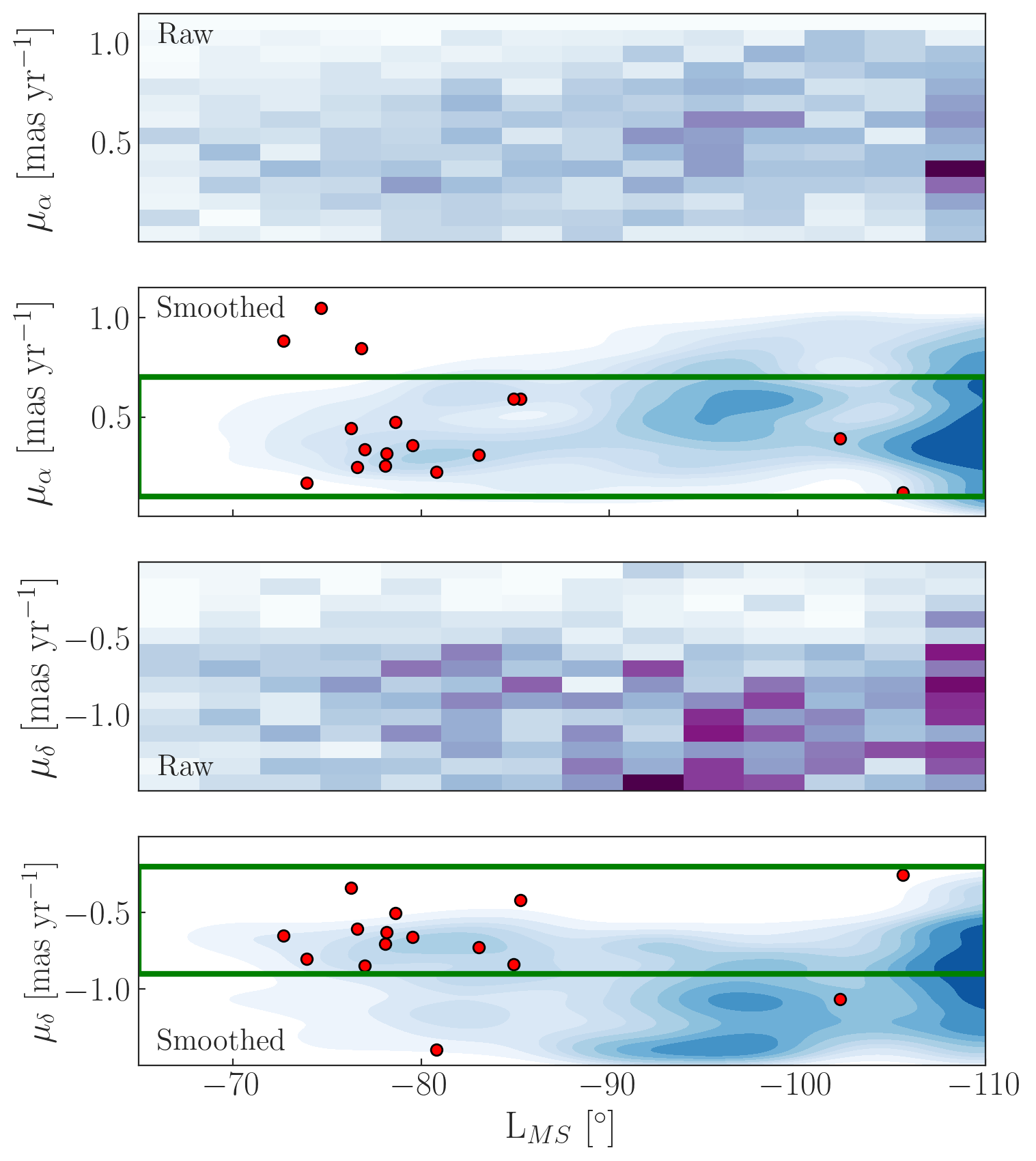}
\caption{The distribution of proper motion values for stars selected using the criteria shown graphically in Figure \ref{fig:cuts} as function of MS longitude. For $\mu_\alpha$  and $\mu_\delta$ we show both the raw binned distribution and the smoothed distribution in blue. The H3-selected stars are superposed on the smoothed {\it Gaia} distributions as red circles. Green horizontal lines show our refined proper motion selection criteria. Stars between the two green lines are retained to produce Figure \ref{fig:mag_streams}.}
\label{fig:final_cuts}
\end{figure}

The portion of the resulting population of 3,967 stars that lies within the plot boundaries is presented in Figure \ref{fig:final_cuts} showing both $\mu_{\alpha}$ and $\mu_{\delta}$ as a function of $L_{MS}$. It is evident that 1) the distribution of stars is not uniformly distributed in proper motion, 2) there is a concentration of {\it Gaia} stars at $-75^\circ \lesssim L_{MS} \lesssim -95^\circ$ that matches the H3 stars to a tighter degree than allowed for by our proper motion selection cuts, and 3) there is a continuing population of stars toward more negative $L_{MS}$ that trends to lower values of $\mu_{\delta}$. 

The selected stars can be divided into two sub-populations. The first is a set that matches that identified initially in H3. The correspondence between the H3-selected stars and these {\it Gaia}-selected stars supports the interpretation of the H3-selected stars as a coherent, physical stellar population and greatly increases the number of such stars. The second is a population that clusters farther along the MS, at more negative $L_{MS}$. This population is difficult to interpret because increasingly negative values of  $L_{MS}$ corresponds to decreasing Galactic latitude, longer sightlines through the halo, and therefore more noise/contamination. Nevertheless, we do not yet dismiss this population entirely. We will return to this population later. 

We use these results to refine our selection of {\it Gaia} stars that correspond to the H3-selected sample. We tighten our proper motion selection to $0.1 < \mu_{\alpha}/({\rm mas\  yr}^{-1}) < 0.7$ and $-0.9 < \mu_{\delta}/({\rm mas\  yr}^{-1}) < -0.2$. In making these cuts we aimed for the following: 1) to retain as many of the H3-selected stars that overlap with the coincident, localized
overdensity of {\it Gaia} stars, and 2) to exclude other overdensities, particularly the one at $L_{MS} \sim -100^\circ$, that do not have corresponding significant populations of H3-selected stars. We retain 891 stars.

\begin{figure*}
\includegraphics[width=\textwidth]{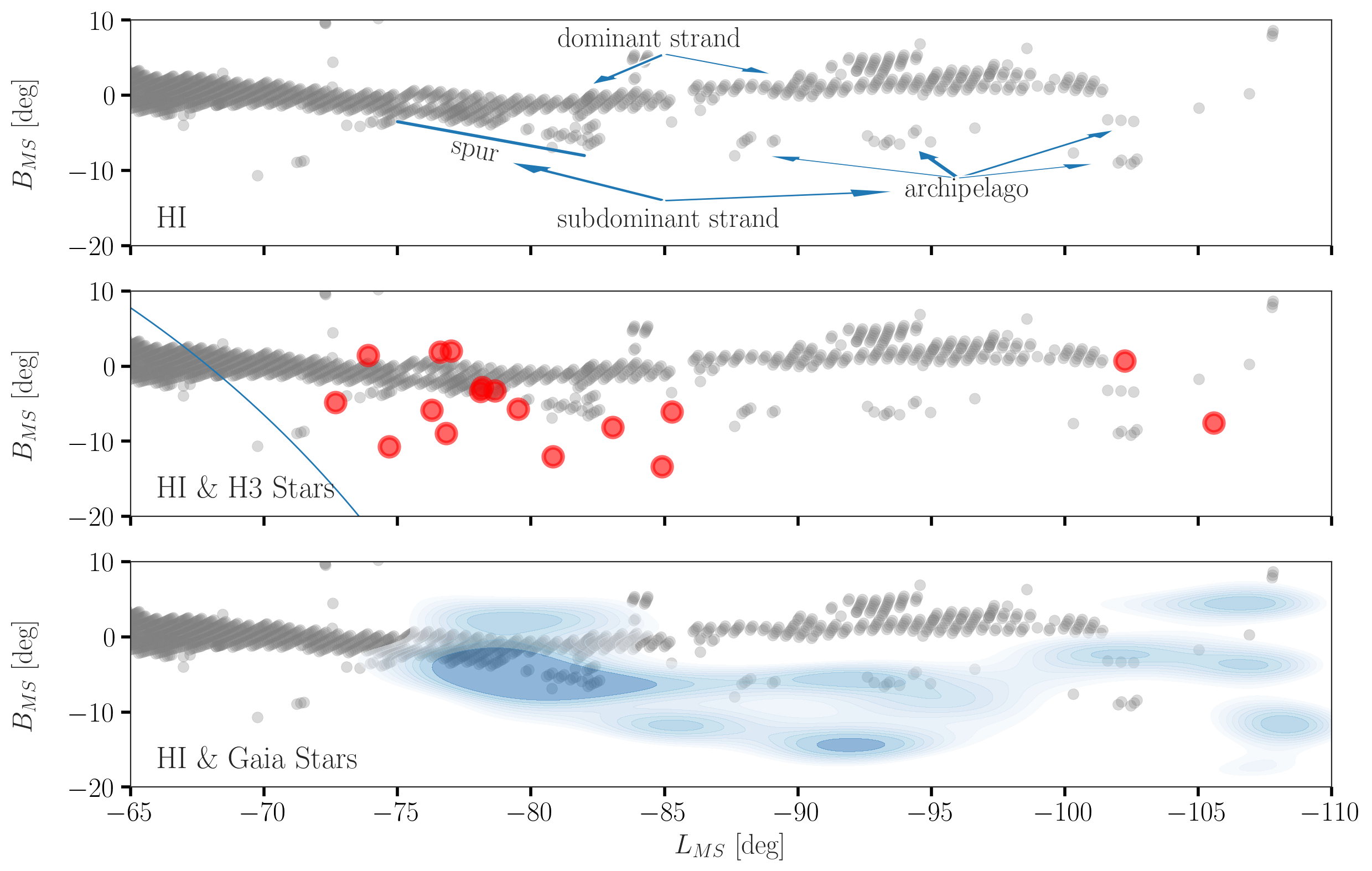}
\caption{Three views of the Magellanic Stream system. Top panel presents the HI distribution along the Magellanic Stream in Stream coordinates \citep{nidever_08}. Note the bifurcation of the Stream at $L_{MS} \sim -75^\circ$ with a small spur to lower $B_{MS}$ that plausibly extends, through a set of small clouds that we name the ``archipelago", out to $L_{MS}\sim -105^\circ$. The data are presented discretely and so the gas, despite presumably being spatially diffuse, is plotted as discrete points. Middle panel adds to this picture of the gaseous Stream the H3-selected stars (red circles) that appear to deviate from the principal Magellanic Stream and follow the spur. The curved line represents the edge of the H3 survey footprint. Bottom panel adds to this picture the {\it Gaia}-selected stars. The distribution of selected stars, smoothed as in Figure \ref{fig:final_cuts}, clearly follows the spur and perhaps, with less certainty, traces the HI archipelago all the way to $L_{MS} \sim -105^\circ$.}
\label{fig:mag_streams}
\end{figure*}

\section{Results}
\label{sec:results}

The populations of gas, H3-selected stars, and our final set of {\it Gaia}-selected stars are presented in Figure \ref{fig:mag_streams}. As noted previously, the MS is a set of apparently intertwined gaseous strands \citep{cohen,morras,putman03,nidever_08}. Because we do not know the distances to the various strands, the 3-D geometry, and therefore the origin of the structures, is unclear.
Nevertheless, in the upper panel of the Figure there appear to be at least two clear strands of the MS 
over the values of $L_{MS}$ included in the Figure. The dominant strand appears to be one that extends from at least $L_{MS}=-65^\circ$ to $L_{MS} \sim -85^\circ$, after which there is a slight gap with the plausible continuation of gas extending from $L_{MS} \sim -86^\circ$ to $L_{MS} \sim -102^\circ$. There is a second strand that bifurcates from the dominant strand at $L_{MS} \sim -75^\circ$, forms first what we refer to as a ``spur" and continues to more negative values of $L_{MS}$, plausibly along an ``archipelago" of H{\small\ I} clouds stretching out as far as the dominant stream but mostly at more negative values of $B_{MS}$ than the dominant Stream (i.e. at $B_{MS} \sim -5^\circ$). We note for completeness, but do not show in this Figure, that the situation is even more complicated because the dominant stream itself shows a velocity discontinuity at the gap \citep{nidever_08}.

In the middle panel of the Figure we compare the neutral gas to the distribution of the H3-selected stars. As suggested by \citetalias{zh3}, the stars appear to be more closely associated with the subdominant MS strand, in particular with the spur to lower $B_{MS}$ at $-83^\circ \lesssim L_{MS} \lesssim -75^\circ$.
In addition, one or both of the H3 stars at $L_{MS}< -100^\circ$ might be part of the extension of this population. As already noted, the small number of stars in the main group of H3-selected stars (15) makes it difficult to reach a definitive conclusion and motivated our study of the {\it Gaia} catalog.

In the lowest panel of the Figure we show the resulting distribution of the {\it Gaia}-selected stars (smoothed kernel density estimation). Our principal result is that we confirm that there is a population of stars that follows the H{\small\ I} spur that was initially associated with the H3-selected stars (i.e. the feature extending over $-83^\circ < L_{MS} < 74^\circ$). It is evident that these {\it Gaia}-selected stars follow the subdominant strand rather than the dominant one. For reference, over the longitude range covered by the 15 H3-selected stars ($-86^\circ < L_{MS} < -74^\circ$ there are 294 {\it Gaia}-selected stars, which alleviates any concern that the H3-selected stars are themselves dominating the feature seen in the {\it Gaia}-selected stars. We shall return to what might appear to some to be a surprisingly close alignment of the stars and gas in \S\ref{sec:discussion}.
Of course, the stars and neutral gas are embedded and surrounded by ionized gas that constitutes the bulk of the gaseous mass removed from the Clouds \citep{fox14,barger}. As such, the presence of stars where there is no H{\small\ I} detected does not mean that there is no associated gas.

\begin{figure}
\includegraphics[width=\columnwidth]{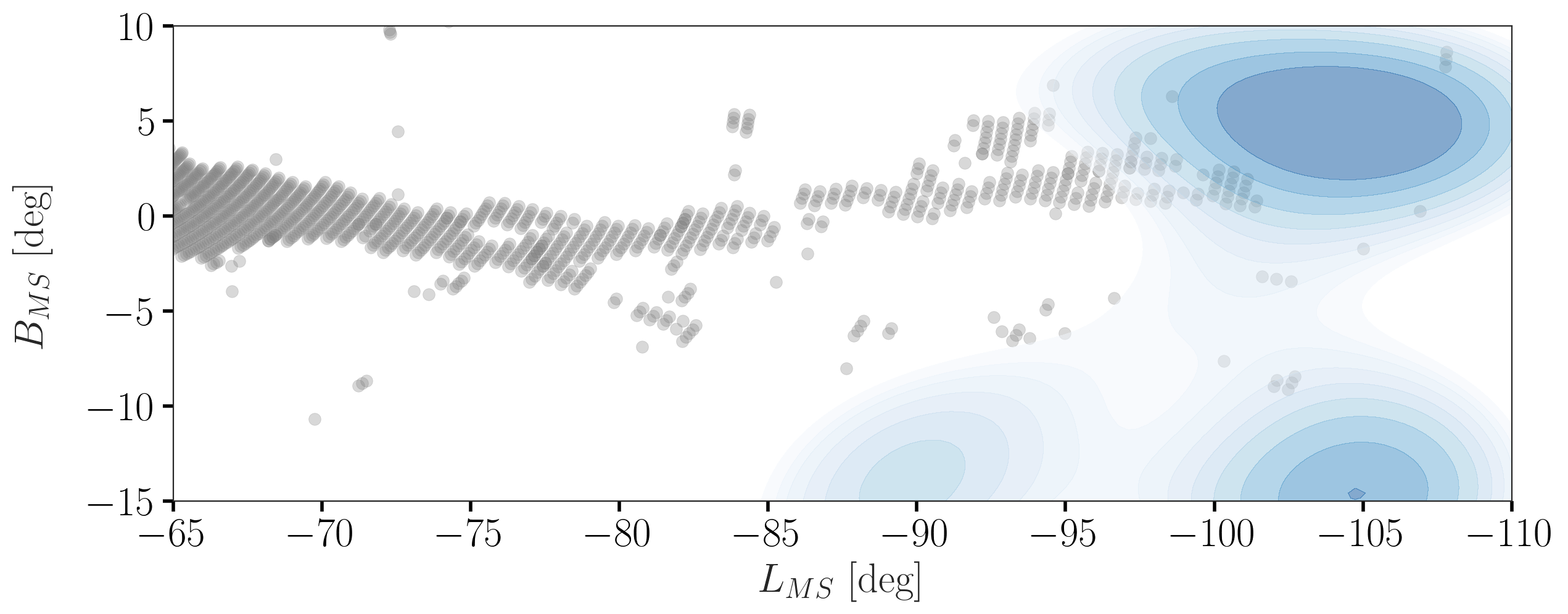}
\caption{The resulting distribution of simulated halo stars selected in the same manner as the {\sl Gaia} stars shown in Figure \ref{fig:mag_streams}. As anticipated, larger fluctuations appear at most negative values of $L_{MS}$. }
\label{fig:gums}
\end{figure}

To provide a control for the detection of the stellar populations,
we apply the same selection cuts to a mock {\sl Gaia} catalog of halo stars with $G<19$ mag created using GUMS \citep{gums}. We present the analogous figure to the lowest panel of Figure \ref{fig:mag_streams} in Figure \ref{fig:gums}. We 
find
large fluctuations in the stellar projected densities at more negative values of $L_{MS}$ because one is approaching the Galactic plane. As a cautionary note, without distance or radial velocity information the largest density fluctuation could be incorrectly identified as a stellar population associated with the tip of the MS. This confusion is not an issue at less negative values of $L_{MS}$, particularly at $L_{MS} \sim -80^\circ$ where we identify our most significant cluster of {\sl Gaia}-selected stars, because the random fluctuations in this region are negligible. Additionally, we remind the reader that for the H3 stars that correspond to this feature we do have velocity information confirming that this is a physically coherent set of stars. 

There are additional concentrations of {\it Gaia}-selected stars that trace the H{\small\ I} archipelago. For these groupings of {\sl Gaia}-selected stars, we stress the tentative nature of that association and the need for follow-up spectroscopy of candidates. Although suggestive, the lack of H3-selected stars along the archipelago prevents us from using kinematics or distances to confirm that these are a physical continuation of the subdominant MS strand. These regions are within the H3 footprint, so a relevant question is why H3 identified at most only two of the stars in these features (i.e., why are 88\% of the H3-selected stars  in the spur rather than in the archipelago). If we define a spur region 
$(-74^\circ < L_{MS} < -86^\circ {\rm \ and\ } -15^\circ < B_{MS} < 5)$ and an archipelago region $(-86^\circ<L_{MS}<-106^\circ {\rm \ and\ } 0^\circ < B_{MS} < -10^\circ)$, we find that 221/392, or 56\%, of the {\sl Gaia}-selected stars are in the spur rather than in the archipelago. The difference between 88\% and 56\% may suggest a higher level of contamination of the {\sl Gaia}-selected sample in the archipelago. As just discussed, we do expect more contamination as we extend to more negative values of L$_{MS}$. Using the mock catalog we find a 3:1 contamination ratio between the archipelago region and the spur. A combination of statistical noise and contamination may explain the difference between the H3- and {\sl Gaia}-selected samples. An alternative explanation is that we are losing H3 stars along the strand because those tend to be at larger distances and hence more challenging for H3. Follow-up spectroscopy of the {\it Gaia}-selected stars in these features is an obvious way forward.

We searched other datasets for existing observations of stars in this area of sky. In particular, we searched among the distance estimates provided by \cite{starhorse} for stars within this region of sky that have 16\% percentile distance estimates that are $>$ 30 kpc. We find 7 matches in the Starhorse LAMOST \citep{lamost} LRS DR7 catalog and 18 in the SDSS DR12 catalog \citep{sdss12}. In neither case do the stars clearly delineate a feature that corresponds to the MS. While there are stars in these samples that are plausibly associated with the {\sl Gaia} features we have identified, we can reach no clear conclusions with these small samples. 

\begin{figure}
\includegraphics[width=\columnwidth]{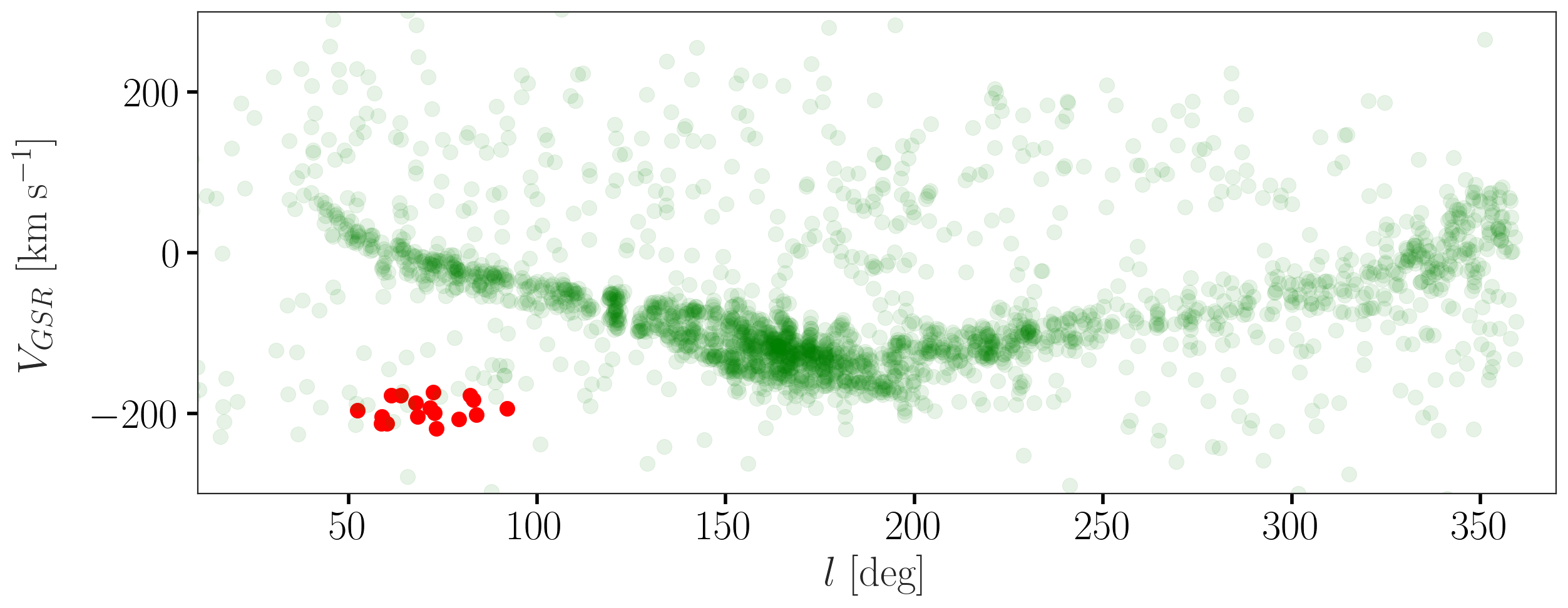}
\caption{The velocity distribution as a function of Galactic latitude for stars  in H3 identified as Sgr members \citep{johnson}, in green, and our H3-selected MS stars, in red. Although the Sgr stream is superposed on our MS stars on the sky, the populations are clearly separable with the additional kinematic information.}
\label{fig:sgr}
\end{figure}

We now return to the question of potential contamination by the Sgr stream. Although the stream crosses our population in projection, in Figure \ref{fig:sgr} we show that the Sgr stream stars in H3 \citep[Sgr\_FLAG=1;][]{johnson} are clearly separate from our population in parameter space. The exception are a few stars from the kinematically hot halo population of Sgr \citep{johnson}. We noted in \S\ref{sec:data} that 5 members of the \citetalias{zh3} sample are now labeled as Sgr members in the H3 catalog and so rejected from our current analysis. If we redo our selection using only labeled Sgr members in H3 we find 6 stars, in other words only one additional star. This set of six is a much more heterogeneous population of stars than our sample. Three of these are well outside the color-magnitude bounds we used for the {\sl Gaia}-selected stars, the stars do not track the H{\small\ I} spur, they have a wide range in [Fe/H] (from $\sim -2.4$ to $-$1.0), and all but one lie outside the proper motion bounds set in Figure \ref{fig:final_cuts}. We conclude that Sgr stars, at least those similar to those identified as such in H3, are not the bulk of our sample and that we have not missed a large number of stream stars due to an incorrect classification as Sgr stars. Likewise, in reference to possible unknown halo stellar overdensities that might contaminate our results, we note that the H3-selected stars are not only well localized on the sky, they align with the MS H{\small\ I}, and perhaps most unlike a putative halo overdensity, have a low internal velocity dispersion, 14 km sec$^{-1}$, and an extreme negative radial velocity, $\sim -200$ km sec$^{-1}$. It is critical to confirm a similar velocity and  low velocity dispersion for the {\sl Gaia}-selected stars.

\begin{figure}
\includegraphics[width=0.9\columnwidth]{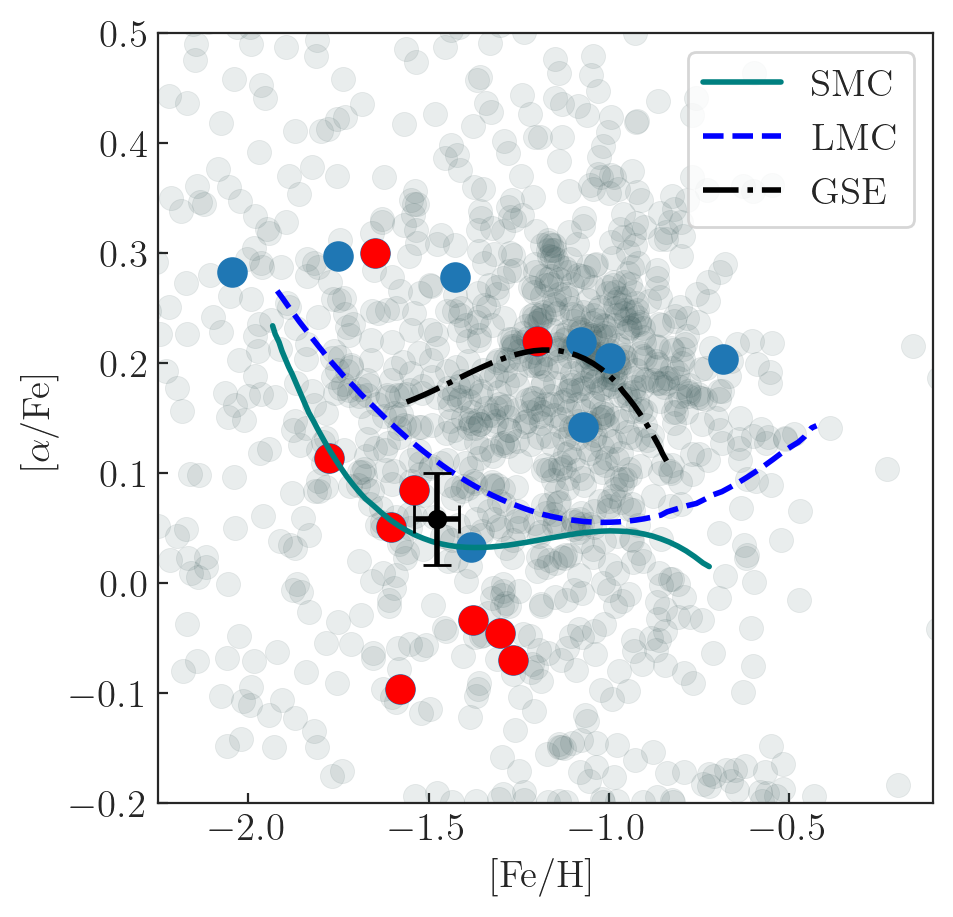}
\caption{Chemical properties of H3-selected stars. Derivation of abundances from H3 spectra are described in \cite{cargile2020}.} We compare the properties of the H3 halo stars (gray), which satisfy all of our selection criteria except the kinematic and MS coordinate ones, to the H3-selected stars. In blue are those H3-selected stars that do not meet the criteria applied to the {\sl Gaia} stars, while in red are those that do. We show tracks in abundance space \citep{hasselquist} for the stellar populations of the SMC, LMC, and Gaia-Sausage-Enceladus (GSE). The point with errorbars indicates the means of the red coded H3-selected stars.
\label{fig:alpha}
\end{figure}

Finally, in closing we return to the H3-selected stars. To even more finely select stars corresponding to the {\sl Gaia}-selected population, we now apply the final proper motion and photometric cuts to the H3-selected stars that we applied when selecting the {\sl Gaia} stars. This results in the rejection of 8 stars from the sample of 17. While this is a large fraction, the selection seems to be doing something sensible (Figure \ref{fig:alpha}). First, it removes the four stars with the largest values of [Fe/H] and these happen to be quite consistent with the peak of the halo distribution in both [Fe/H] and [$\alpha$/Fe]. Second, it removes 7 of 9 of the stars with the highest values of [$\alpha$/Fe], again values that are more consistent with a halo population. Although the trimming may be overly aggressive because we are applying our tightest proper motion cuts and because color and metallicity are dependent, this subsample is the closest H3 analog to the {\sl Gaia}-selected population. 

The remaining stars have $\langle$[Fe/H]$\rangle = -1.48 \pm 0.06$ and $\langle$[$\alpha$/Fe]$\rangle = 0.06 \pm 0.04$, which are slightly more consistent with the previously measured properties of SMC stars than of LMC stars (Figure \ref{fig:alpha}). The discriminatory power of this measurement is marginal because of the level of uncertainty shown in Figure \ref{fig:alpha}, but also because there is likely internal scatter within each galaxy, because  the properties of extracted stars may not reflect the bulk properties, and because there are slight systematic differences among abundance derivations among studies \citep[cf.][]{cargile2020}. Furthermore, the \cite{pardy} models (their Figure 4) suggest that there could be both SMC and LMC material in this region of the MS. Their models, which are for the gas component, suggest that while SMC material dominates in this region, there is an underlying population of LMC material as well. Such a mix will further complicate the interpretation of Figure \ref{fig:alpha}. Alternatively, the stars could also come from an unknown progenitor, presumably a completely disrupted satellite of the Clouds, with chemical properties comparable to those of the SMC.

\section{Discussion}
\label{sec:discussion}

Before associating the stellar populations we have identified here with specific MS features, we briefly review the status of field. \cite{nidever_08} presented the most complete case for the presence two H{\small\ I} strands in the MS, although  evidence for such existed for over forty years \cite{cohen,morras,putman03}. \cite{nidever_08} focused on the two strands for $L_{MS}>-40^\circ$ (toward the LMC and SMC) in an attempt to trace the strands back to their source. For one strand, they make a convincing case of an LMC origin. For the other, they are unable to distinguish between and LMC, SMC, or Bridge as the possible source. \cite{fox13}, using absorption line spectroscopcy, confirmed, based on metallicity arguments, the LMC origin scenario proposed by \cite{nidever_08} for the one strand, but conclude from most of their sight-lines that much of MS gas originated in the SMC.
It is admittedly difficult to unambiguously connect the various uneven H{\small\ I} strands from where we identify two strands to where \cite{nidever_08} do. What we refer to as the dominant strand is not necessarily the same as what appears to be the dominant strand closer to the Clouds.

There are two key aspects of our results. First, we have identified a Gaia population of stars that corresponds to the H3-selected population. This result is independent of any association with the gaseous MS and confirms the existence of a stellar MS component at moderate distances ($\sim 30-60$ kpc). Second, there is an intriguingly close, detailed association between the projected location of the Gaia-selected stars and what is the subdominant strand of the H{\small\ I} MS at these values of L$_{MS}$.  This aspect is somewhat more complex to interpret given the expectation that stars and gas will be offset by hydrodynamical forces acting on the gas. 

\begin{figure}
\includegraphics[width=\columnwidth]{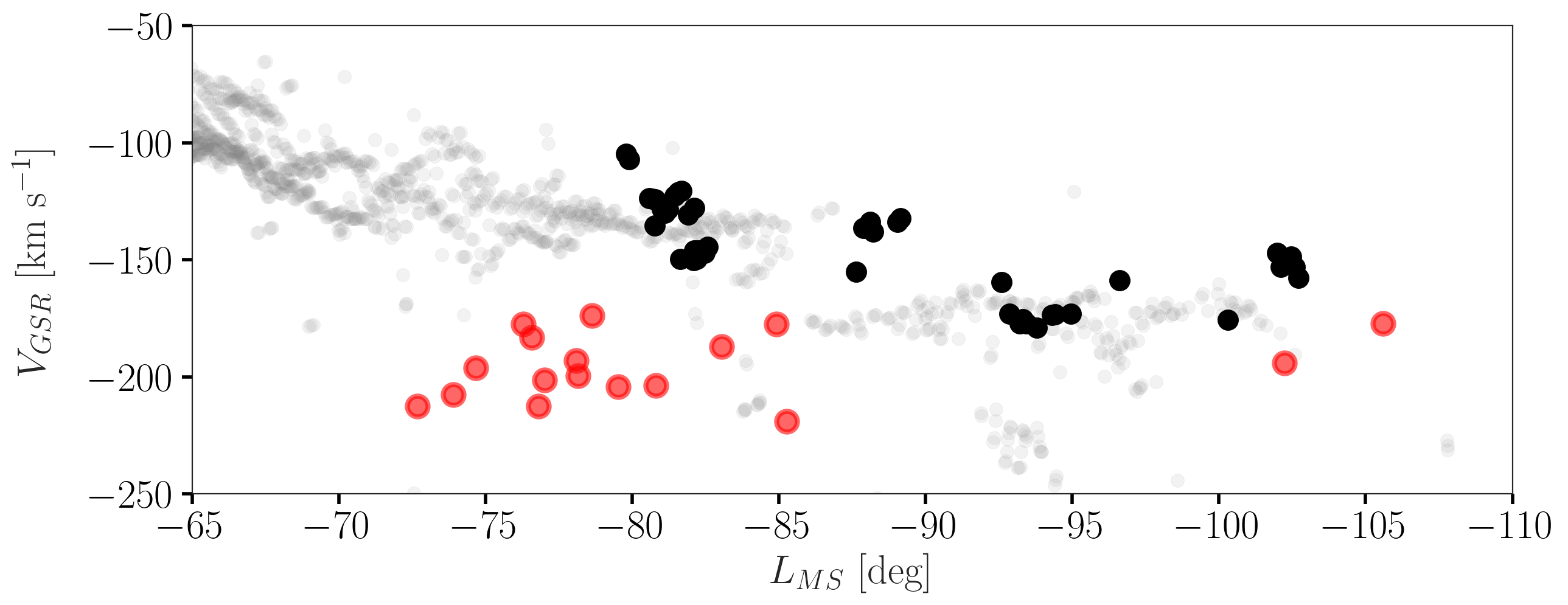}
\caption{Kinematics of MS components. The HI, shown in grey, \citep{nidever_08}, is compared both to the HI in the subdominant strand (defined for this purpose as lying within $-77^\circ < L_{MS} < -110^\circ$ and $-15^\circ <B_{MS}<-4^\circ$), shown in black, and the H3-selected stars, shown in red. }
\label{fig:vel_comp}
\end{figure}

Drag produced by a galactic corona has been invoked to explain offsets in stellar and gaseous tails more broadly \citep{hibbard} and shown to produce offsets among stars and gas in the leading arm of the MS \citep{price-whelan,nidever+19}. If the apparent association is physical and there is drag on the gas, then the resulting displacement must be principally along the line of sight in those instances where stars and gas appear coincident in Figure \ref{fig:mag_streams}. This supposition has two implications: 1) the motion of both the stars and the gas we are considering must be principally along the line of sight and 2) there should be a radial velocity displacement between the gas and the stars. Regarding the first point, the H3-selected stars have radial velocities at the extreme of the distribution for their distance (Figure \ref{fig:selection}) and, as such, if they are bound to the Milky Way they cannot have a comparably large transverse velocity component. Regarding the second point, the radial velocities of the H3-selected stars and the H{\small\ I} are indeed offset (Figure \ref{fig:vel_comp}) in the expected sense (the gas is infalling more slowly). Given the close association between stars and gas in Figure \ref{fig:mag_streams} and these arguments, we prefer an interpretation where the association is physical, but acknowledge that the association may instead be fortuitous and ephemeral.

\begin{figure}
\includegraphics[width=\columnwidth]{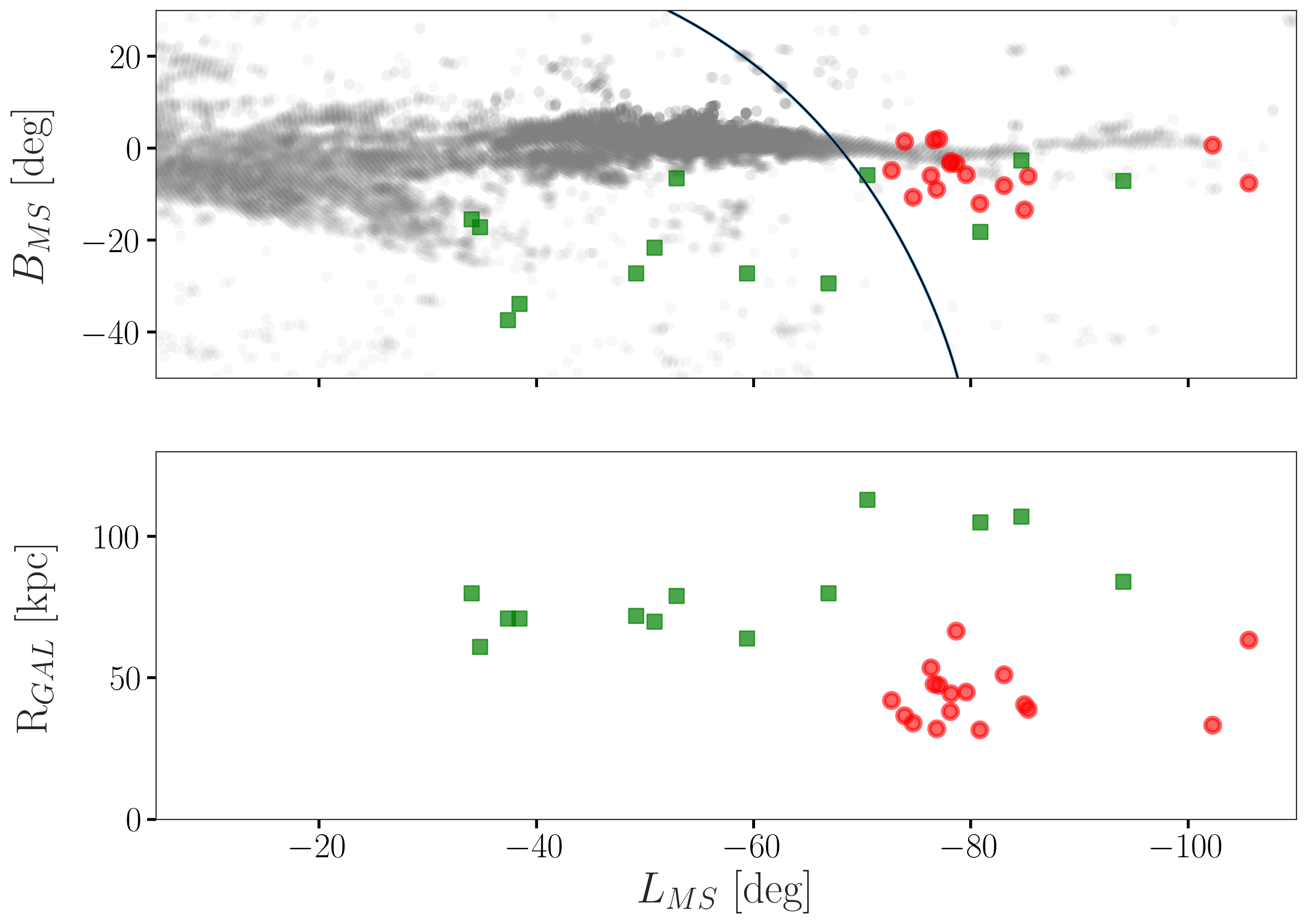}
\caption{Comparison of our set of stars to those identified by \citetalias{chandra}. The \citetalias{chandra} stars (green squares) are distributed over a far larger area of sky than our stars (red circles) and so only three or four overlap our search area. Those that are projected near our set of stars are at about a factor of 2 farther away and therefore likely trace a different physical structure. They were also selected to have different angular momenta. We plot the H{\small\ I} distribution in the upper panel from \cite{nidever_08} and a curve showing the boundary of the H3 footprint for reference.}
\label{fig:chandra}
\end{figure}

The nature of the dominant and subdominant MS strands as we have defined them is fundamentally different. 
The lack of stars from our dataset in the dominant strand --- drawn either from H3 or {\it Gaia} --- does not necessarily indicate that it does not contain stars. Both surveys are increasingly insensitive to stars at larger distances. The stars identified by \citetalias{chandra} lie between $60-120$~kpc, and some may be associated with the dominant strand (Figure \ref{fig:chandra}). If so,  then the dominant and subdominant MS strands as we have identified them are separated by tens of kpc ($\sim 100$ kpc vs. 47 kpc) and are only closely aligned in projection. Such overlapping streams have been identified in simulations, either due to the overlapping of two separate tidal tails or the orbital wrapping of a single tail \citep{diaz}. Models of the MS show gas originating in both the LMC and SMC \citep{pardy}.

An alternative possibility is that the dominant strand is devoid of stars and arises either from tidally stripped gas that was originally at a radius beyond where there were many stars \citep{diaz} or from gas that was blown out of the central galaxy and then stripped or tidally removed  \citep{nidever_08}. In either version of this alternative scenario, the \citetalias{chandra} stars are unrelated to this dominant strand of the MS and are likely tidal stellar debris drawn from the extremities of the Clouds. Such stellar halo-like components have been traced in the periphery of the Clouds \citep{belokurov,ely,gaia-mc} and likely have a complicated history \citep[e.g.,][]{massana24}. 

The subdominant MS strand investigated in the present work shows a close correspondence between stars and gas in the detailed density structure along the  strand. The presence of stars confirms that this gaseous strand is a tidal feature. The correspondence between gas and stars suggests that material originated from a galactic  disk, which would contain both gas and stars. The metallicity and $\alpha$ enrichment of our H3-selected stars is consistent with an SMC origin.  In support of such an interpretation, we note that absorption line studies of the MS \citep{fox13} and numerical simulations \citep{pardy} suggest that much of the MS at these values of L$_{MS}$ originates from the SMC.
Nevertheless, more complex scenarios cannot yet be definitively excluded.

To expand on this point, consider that in $\S$\ref{sec:results} we argue that the stellar population within the subdominant strand of the MS does not seem related to Sgr stream stars in the H3 dataset. 
However, it remains plausible that an additional, hitherto undetected component of the Sgr stream could contribute stars matching the population presented here. 
In particular, the simulations of \cite{Vasiliev2021} --- which are constrained to match observations of the main stream body --- predict early-stripped debris that are similarly offset from the stream track as the stars considered here. Although we consider this specific scenario to have low probability given the correspondence of our identified stars with the MS gas, the limited footprint of the (northern) H3 Survey prevents us from investigating whether the population studied here extends closer to the Sgr stream (see Figure~\ref{fig:chandra}). 
Forthcoming surveys --- chiefly the fifth-generation Sloan Digital Sky Survey from the southern hemisphere (SDSS-V; \citealt{Kollmeier2017}) and 4MOST  \citep{4most} --- should be capable of resolving this question. 

\section{Conclusions}
\label{sec:conclusions}

Using an H3-selected sample of 17 stars that correspond to the population identified by \citetalias{zh3} as a potential stellar counterpart to the gaseous Magellanic Stream (MS), we selected 891 stars from the {\it Gaia} catalog to expand the number of known stars in this component and to map this population. We find that the MS consists of at least two different strands. The dominant one, as defined by gas content, for $-65^\circ < L_{MS} < -110^\circ$, is devoid of stars with Galactocentric distances $\lesssim 55$ kpc and may be traced by stars at larger distances (\citetalias{chandra}). The subdominant one shows a close correspondence between stars and gas and lies at distances $\lesssim 55$ kpc. This finding demonstrates conclusively that this feature, at least, is tidal in origin. Given the association between gas and stars, and the mean metallicity of the stars, we suggest that it is tidal material drawn from the body of the SMC, but this finding is marginally significant and requires additional confirmation. Of course, more surprises  \citep[e.g.][]{nidever24} are likely to come in the study of the intriguing, complex Magellanic system.

\acknowledgments
\label{sec:acknowledgments}

We thank Elena D'Onghia, \'Oscar
Jim\'enez-Arranz, Nikolay Kacharov, Denis Leahy, David Nidever, Elena Sabbi, Katime Santrich, Jianling Wang, Yanbin Yang, and an anonymous referee for valuable comments that improved this paper.
We thank MMTO staff and the CfA and U. Arizona TACs for their support of the H3 Survey. Observations reported here were obtained at the MMT Observatory, a joint facility of the Smithsonian Institution and the University of Arizona. This work has made use of data from the European Space Agency (ESA) mission
{\it Gaia} (\url{https://www.cosmos.esa.int/gaia}), processed by the {\it Gaia}
Data Processing and Analysis Consortium (DPAC,
\url{https://www.cosmos.esa.int/web/gaia/dpac/consortium}). Funding for the DPAC
has been provided by national institutions, in particular the institutions
participating in the {\it Gaia} Multilateral Agreement.

\vspace{5mm}
\noindent
{\bf Facilities}: {MMT (Hectochelle), \emph{Gaia}}

\vspace{5mm}
\noindent
{\bf Software: }{\texttt{IPython \citep{Perez2007}, matplotlib \citep{Hunter2007}, numpy \citep{van2011numpy}, Astropy \citep{astropy:2018}, SciPy \citep{2020SciPy-NMeth},
Gala \citep{gala}, galpy \citep{bovy}}}

\bibliography{bibliography}{}
\bibliographystyle{aasjournal}

\end{document}